\documentstyle[epsf,emulateapj]{article}

\received{25 June 1996}
\revised{5 November 1996}

\slugcomment{Astrophysical Journal, in press}

\lefthead{SARAZIN \& MCNAMARA}
\righthead{X-RAY OBSERVATIONS OF A2597}

\begin{document}

\title{ROSAT X-RAY OBSERVATIONS OF THE COOLING FLOW CLUSTER A2597}

\author{Craig L. Sarazin}
\affil{Department of Astronomy, University of Virginia, \\
P.O. Box 3818, Charlottesville, VA 22903-0818; \\
cls7i@virginia.edu}

\and

\author{Brian R. McNamara}
\affil{Harvard-Smithsonian Center for Astrophysics, \\
60 Garden Street, Cambridge, MA 02138; \\
brm@cfa241.harvard.edu}

\begin{abstract}
The cluster A2597 was observed in X-rays with
the ROSAT PSPC and HRI detectors.
The X-ray emission from the cluster extends out to a radius of at least
2.37 Mpc.
The X-ray isophotes are oriented similarly to the optical isophotes
of the central cD galaxy and to the isopleths of the
galaxy distribution in the cluster, but are otherwise quite
regular, suggesting that this cluster is reasonably relaxed and
in hydrostatic equilibrium.
The merged HRI and PSPC surface brightness profile is not adequately
fit by a beta model because of the central X-ray surface brightness peak,
indicating the presence of a cooling flow.
If the central 108 arcsec in radius are excluded, an acceptable fit is
found which gives $\beta = 0.64^{+0.08}_{-0.03}$
but only an upper limit to the core radius,
$r_{core} < 78$ arcsec.
The gas density and accumulated mass were derived as a function of
radius, and the hydrostatic equilibrium condition was used to determine the
total gravitational mass, also as a function of radius.
Within a radius of 2 Mpc, we found masses of
$M_{gas} = 1.2 \times 10^{14}$ $M_\odot$ and 
$M_{tot} = 6.5 \times 10^{14}$ $M_\odot$, and
a gas mass fraction of about 19\%.
However, our poor knowledge of the spatial variation of the gas
temperature makes the total mass values uncertain by at least a factor
of two.

The ROSAT PSPC X-ray spectrum of the cluster was determined.
Both the overall cluster spectrum and the spatially resolved spectra
within 300 kpc require the presence of both hot gas and a cooling flow
in the spectrum.
The spectrally determined total cooling rate of
$\dot{M} = 344^{+75}_{-67} \, M_\odot$ yr$^{-1}$ is in good
agreement with those derived from analyses of the X-ray surface
brightness profile from the $Einstein$ IPC
and the ROSAT HRI images.
The ROSAT spatially resolved X-ray spectra indicate that the cooling component
is distributed over the inner $\sim$300 kpc in radius of the cluster.
The presence of the cooling flow in the central regions and poorer
statistics in the outer regions prevent us from deriving an
accurate profile for the variation of the ambient cluster gas temperature.

We do not detect any significant excess X-ray absorption toward the center
of A2597, and we set a very conservative upper limit on the
excess column in front of the cooling flow region of
$\Delta N_H < 1.72 \times 10^{20}$ cm$^{-2}$.
A2597 is one of the few cooling flows toward which
a large column has been detected in radio observations
(O'Dea et al.\ 1994a).
A total column of
$N_H \approx 1 \times 10^{21}$ cm$^{-2}$ was seen in 21 cm
absorption against the very small central radio source.
Our X-ray upper limit is not inconsistent with the 21 cm detection
if the absorber only covers the small region ($\sim$6 arcsec) occupied
by the radio source.
However, we can rule out an absorber with a uniform column which covers
the entire cooling flow, or even just the smaller region of the
extended optical emission line nebula.
\end{abstract}

\keywords{cooling flows ---
galaxies: clusters: general ---
galaxies: clusters: individual (A2597) ---
galaxies: cD ---
intergalactic medium ---
X-rays: galaxies
}

\section{INTRODUCTION} \label{sec:intro}

The Abell richness class 0 cluster A2597
(Abell, Corwin, \& Olowin 1989)
at a redshift of $z = 0.0852$ is elliptical but quite regular
in its galaxy and gas distributions (e.g., Tr\`evese et al.\ 1992b;
Buote \& Tsai 1996),
with an elongated cD galaxy at its center
(McNamara \& O'Connell 1993).
This central galaxy in A2597 contains the moderately luminous
($P_{20} = 5.98 \times 10^{25}$ W/Hz)
radio source PKS~2322-122
(Owen, White, \& Burns 1992;
Ball, Burns, \& Loken 1993).
Recent radio observations show that this very small source
($\la$6 arcsec in diameter) has emission from a radio core,
a strongly bent jet, and two steep spectrum radio lobes on either
side of the nucleus
(Sarazin et al.\ 1995b, hereafter Paper I).

A2597 is a powerful X-ray source
with a total luminosity of $L_X = 6.45 \times 10^{44}$ erg
s$^{-1}$
(2--10 keV;
David et al.\ 1993).
The cluster contains a strong central cooling flow
(Crawford et al.\ 1989; Paper I),
with a total cooling rate of
$\dot M \approx $ 300--400 $M_\odot$ yr$^{-1}$
out to a radius of
$r_c \approx 200 $ kpc.
As is often true of strong cooling flows, the
central cD galaxy in A2597 has extensive optical and UV line
emitting filaments
(Hu 1988;
Heckman et al.\ 1989;
Crawford \& Fabian 1992).
The X-ray emission in the central regions is elongated along
a direction which it perpendicular to the orientation of the
radio lobes
(Paper I).
Similar X-ray---radio anti-correlations have been seen in a
number of other cooling flow clusters
(e.g., B\"ohringer et al.\ 1993;
Harris, Carilli, \& Perley 1994;
Sarazin, Baum, \& O'Dea 1995a).

As is also true of the central galaxies in many other cooling
flow clusters, the central optical colors of the cD galaxy in
A2597 are considerably bluer than a typical giant elliptical
galaxy
(Crawford \& Fabian 1992;
McNamara \& O'Connell 1993).
The bluest regions lie in lobe-like structures which extend
radially $\sim 5-7$ kpc
on either side of the I-band nucleus with a NE/SW orientation
(McNamara \& O'Connell 1993).
These blue optical lobes appear to be oriented along the
axis of the radio jets in the central galaxy
(Paper I).
The optical lobes could originate from a number of processes.
However,
the most plausible on morphological and energetic grounds are
radio-lobe-induced
star formation
or scattered nonthermal radiation beamed anisotropically from the
nucleus
(e.g., Paper I).
There is also evidence for the presence of dust in the central
regions from the optical color distribution in the cD
galaxy and from the ratio of H$\alpha$/Ly$\alpha$ line emission
(Hu 1992;
McNamara \& O'Connell 1993).

Through a re-analysis of $Einstein$ Solid State Spectrometer
X-ray spectra of the central regions of cooling flow clusters,
White et al.\ (1991) found evidence for very large amounts of
excess soft X-ray absorption.
A2597 was not included in the sample studied by White et al.
However, for strong cooling flows like A2597, the excess columns
were typically $\ga 1 \times 10^{21}$ cm$^{-2}$.
ROSAT PSPC and ASCA spectra have confirmed this absorption
in a number of clusters
(Allen et al.\ 1993;
Fabian et al.\ 1994),
and the ROSAT PSPC spectral images have shown that the excess
absorption is confined to the inner cooling regions of the
cluster
in several cases
(Allen et al.\ 1993;
Irwin \& Sarazin 1995).
In general,
the cold material producing this absorption has not been
detected at non--X-ray wavelengths, despite considerable
efforts
(e.g., McNamara \& Jaffe 1993;
Antonucci \& Barvainis 1994;
O'Dea et al.\ 1994b;
Voit \& Donahue 1995).

A2597 is an exception to this.
O'Dea et al.\ (1994a) detected 21 cm H~I absorption
toward the central radio source in the cD galaxy at the center of
the cooling flow.
The total column is $N_H \approx 1 \times 10^{21}$ cm$^{-2}$,
which is comparable to the columns derived from X-ray spectra
of other cluster cooling flows.
The absorber has a component which is extended at least over
the rather small diameter of the radio source ($\sim$6 arcsec).
Unfortunately, there is no X-ray--derived absorbing column for
this cluster with which to compare the 21 cm detection.
The Galactic column toward the cluster is relatively small
($N_H = 2.45 \times 10^{20}$ cm$^{-2}$; Stark et al.\ 1992),
making the detection of a large column of excess X-ray absorption,
if present, straightforward.

We have observed A2597 with both the ROSAT High Resolution Imager
(HRI) and Position Sensitive Proportional Counter (PSPC).
The HRI results on the inner structure of the cooling flow
and its relation to the radio source have already been
presented in Paper I.
In the present paper, we will concentrate on the results from the
ROSAT PSPC.
The purpose of these observations was to use the high
sensitivity of the PSPC to map the large-scale X-ray emission
of A2597 and to use the spectral resolution to determine the
thermal structure of the cluster.
In addition, the PSPC is an ideal instrument for searching
for excess soft X-ray absorption, which we can compare to the
radio
detection of neutral hydrogen and to the excess absorption
measurement toward other cooling flow clusters.
The X-ray observations are described in
\S~\ref{sec:observations}.
The overall X-ray image is discussed in \S~\ref{sec:images}.
The X-ray spectrum of the cluster is analyzed in
\S~\ref{sec:spectra},
and the temperature structure is presented in
\S\S~\ref{ssec:spec_whole} and \ref{ssec:spec_annuli}.
We derive an upper limit on the excess X-ray absorption and
compare
to the radio observations in
\S~\ref{ssec:spec_excess}.
The shape of the X-ray emission and its radial profile are
determined
in \S\S~\ref{ssec:ellipse} and \ref{ssec:surf_bright}.
These are used to determine the distribution of the gaseous and
total gravitating mass in \S~\ref{sec:masses}.
The results are summarized in
\S~\ref{sec:conclusion}.
Other X-ray sources in the field are listed in an Appendix.

All distance-dependent values in this paper assume
$H_o = 50$ km s$^{-1}$ Mpc$^{-1}$ and $q_o = 0.0$.

\section{X-RAY OBSERVATIONS} \label{sec:observations}

The cluster was observed for 7243 seconds with the ROSAT
Position Sensitive Proportional Counter (PSPC) on 27 November,
1991
(all of the final results presented here are based on the Rev.~2
version of the data) and
with the ROSAT High Resolution Imager (HRI) for a
total of 17,811 seconds.
The HRI observation was done in two intervals:
6-8 December, 1991 and the second was 6-9 June, 1992.
The structure of the cooling flow as seen in the ROSAT HRI data
was discussed previously in Paper I,
where the X-ray structure was compared to radio and optical
structures
at the cluster center.
Here, we will concentrate on the PSPC data.
We will only discuss the global properties of the HRI data which
were not addressed in Paper I.

The PSPC data was screened for periods of high background based on
a Master Veto Rate $>$ 170
(Plucinsky et al.\ 1993),
for other times of high background,
for periods of 15 seconds after switching to the high voltage,
and for periods with an uncertain aspect solution.
The analysis was done with a combination of the IRAF/PROS
software,
the XSPEC/XANADU software, the XSELECT program, and Snowden's
SXRB package for the analysis of diffuse X-ray emission
(Snowden 1995).
The screened data contained a total exposure of 6721 seconds
and the average Master Veto Rate was 87.25.
The data were corrected for particle background, scattered
solar X-ray background, the long term enhancements (LTE) in this
background, and after-pulses.
With these backgrounds removed, the image was corrected for
exposure and vignetting.

\begin{figure*}[htb]
\vskip4truein
\includegraphics{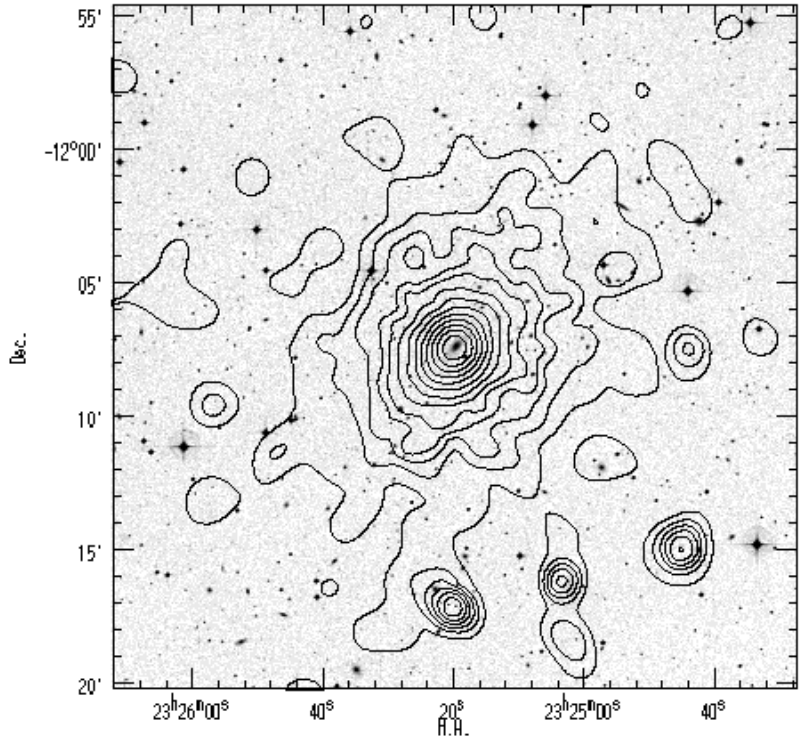}
\caption{A contour plot of the X-ray surface brightness of
the central $34 \time 34$ arcmin ($4.4 \times 4.4$ Mpc)
A2597 cluster.
The X-ray contours are superposed on the optical image of the
cluster
from the Digitized Sky Survey.
The contours are logarithmically spaced with five contours per
dex,
and the lowest contour corresponds to 0.0014 cts sec$^{-1}$
arcmin$^{-2}$
in the 0.4 -- 2.4 keV ROSAT band (PI channels 42--247).
The image has been corrected for background, vignetting, and
exposure,
and adaptively smoothed to a signal-to-noise of five per
smoothing
beam.
The coordinates are J2000.}
\label{fig:xray_cluster}
\end{figure*}

The Appendix gives the image of the whole field of view of
the PSPC observation (Figure~\ref{fig:x-ray_whole} below), and a list
of the field X-ray sources.
We attempted to use optical identifications of these sources
to check and improve the positions in the X-ray image.
Unfortunately, the optical identifications are uncertain and do
not lead to a simple, consistent transformation of the ROSAT
coordinates which would improve the correspondence the optical coordinates.
Similarly, the differences between the ROSAT HRI and PSPC positions
of the points sources in common do not suggest a simple
transformation of the HRI and PSPC coordinates to bring them
into better agreement.
In what follows, we will assume that the peak in the X-ray
surface
brightness of the A2597 cluster agrees with the nucleus of the
central cD galaxy as determined from radio observations,
R.A.\ = 23$^{\rm h}$25$^{\rm m}$19{\fs}64 and
Dec.\ = -12{\arcdeg}07{\arcmin}27{\farcs}4
(Paper I).

\section{X-RAY IMAGE} \label{sec:images}

Figure~\ref{fig:xray_cluster} shows the image of the X-ray
emission from the central $34 \time 34$ arcmin ($4.4 \times 4.4$ Mpc)
of the cluster, superposed on the optical image of the cluster
from the Digitized Sky Survey.
The X-ray image was cleaned, background subtracted, exposure and
vignetting
corrected, and  adaptively smoothed as discussed above.
This image is restricted to the hard band on ROSAT
of 0.4 -- 2.4 keV (PI channels 42--247).
The cluster X-ray emission is quite hard, particularly away from
the bright cooling flow region at the center.
Restricting the emission to the hard band removes much of the
background and emission by other sources, but does not 
affect the number of photons from the cluster as strongly.

\begin{figure*}[htb]
\vskip4truein
\includegraphics{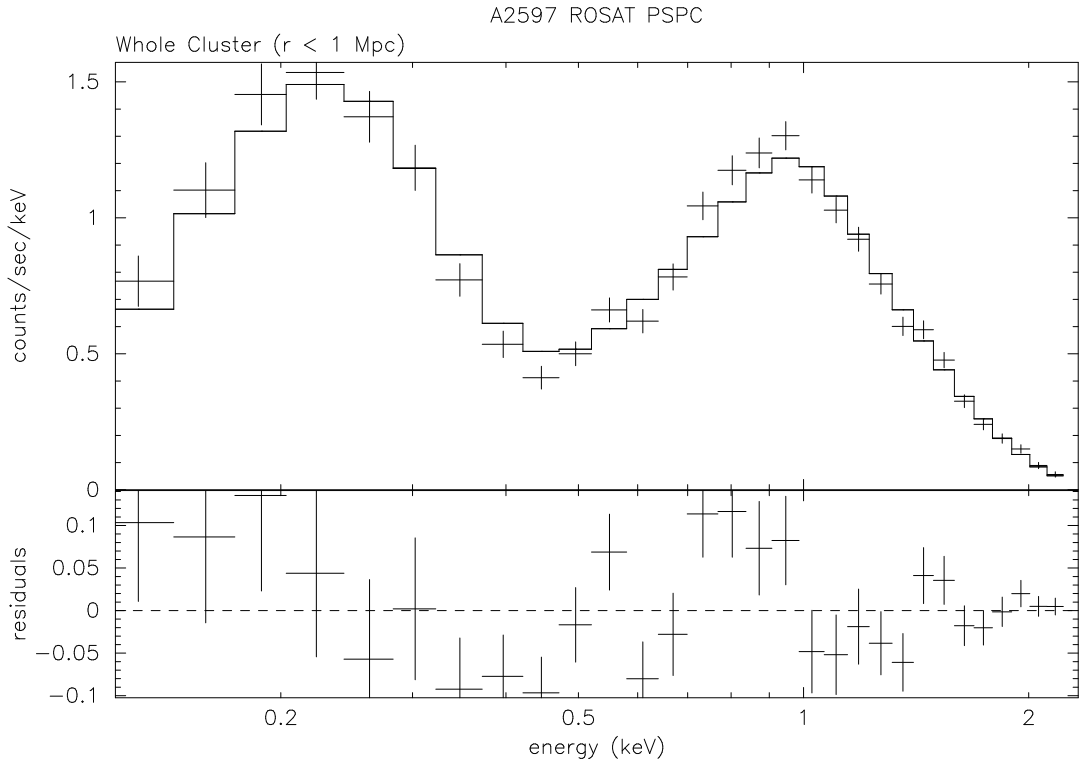}
\caption{The ROSAT PSPC X-ray spectrum for the whole A2597
cluster ($r \le 1$ Mpc) as a function of the measured photon
energy is shown is the upper panel.
The crosses give the data points with 1-$\sigma$ error bars,
while the histogram is the best-fit single temperature model
(Table~\protect\ref{tab:spectra}).
The width of the data points or histogram steps is the width of
the energy channels used to accumulate the data.
The lower panel gives the residuals to the fit (in counts/sec/keV).}
\label{fig:whole_spectrum}
\end{figure*}

\section{SPECTRAL ANALYSIS} \label{sec:spectra}

\subsection{Global Spectra} \label{ssec:spec_whole}

The spectral resolution of the ROSAT PSPC allows the
determination of the
spectral properties of clusters,
at least when the temperature is low or the observation
contains a sufficiently large number of photons.
The only published spectral information on A2597 is from the
Monitor Proportional Counter (MPC) on the $Einstein$ Observatory,
which provided a best-fit temperature of 9.1 keV with a
90\% confidence lower limit of 3.8 keV
(David et al.\ 1993).
We analyzed the PSPC spectrum of A2597 using the facilities
within PROS and XSPEC.
We present the results obtained with XSPEC;
the results of spectral fitting with PROS were generally consistent.
Only counts in the photon energy range 0.11 -- 2.1 keV were used
(PI channels 3--32).
The two higher energy channels (PI channels 33 and 34) have
uncertain calibrations and were excluded.
All fitted spectral bins had $>$10 counts.
The errors in the counts were determined from the square root of
the number of photons;
we did not use the PROS Poisson errors because these
often result in unrealistically low values of the $\chi^2$
statistic.
Some of the results of the spectral fitting are summarized in
Table~\ref{tab:spectra}.
All of the errors or upper limits given for spectral parameters
are at the 90\% confidence level.

First, we determined the global spectrum of the cluster
by extracting the counts from a circle of radius
1 Mpc (7.76 arcmin).
The spectrum was corrected for particle and X-ray background
using a circular annulus extending from 7.76 to 13.4 arcmin.
The initial model we adopted for the spectrum was a Raymond-Smith
model
(Raymond \& Smith 1977)
for the thermal emission and the Wisconsin model
(Morrison \& McCammon 1983)
for foreground Galactic absorption.
We refer to this as a ``single temperature'' model in
Table~\ref{tab:spectra} and in the following discussion.
The best-fit single temperature model for the whole cluster
spectrum had
a temperature of $kT = 2.61^{+0.68}_{-0.48}$ keV, a hydrogen
column density of $N_H = 1.89^{+0.21}_{-0.20}$ cm$^{-2}$, and
a cosmic heavy element abundance fraction of $0.53^{+0.34}_{-0.21}$.
Figure~\ref{fig:whole_spectrum} shows the observed X-ray
spectrum, the best-fit single temperature model, and the residuals
to the fit.
This is a poor fit, with a $\chi^2$ of
47.88 for 26 degrees of freedom (d.o.f.).

The measured Galactic hydrogen column toward A2597 is
$N_H = 2.45 \times 10^{20}$ cm$^{-2}$
(Stark et al.\ 1992).
This is slightly larger than the column
determined from the PSPC spectrum.
In the region around A2597, the Galactic columns varies by about
$ 0.2 \times 10^{20}$ cm$^{-2}$, and it is likely that there are
additional systematic uncertainties at least this large both in the
determination of the Galactic column and 
in the X-ray determination of the column density
(for example, PROS and XSPEC give slightly differing values.)
Thus, the X-ray and radio determinations of the Galactic column
are in reasonable agreement if systematic errors are included.
The heavy element abundance is poorly determined, but is similar
to the values generally found for X-ray clusters.

The best-fit temperature is incompatible
with the temperature derived from the $Einstein$ MPC.
David et al.\ (1993) found a best-fit temperature of
9.1 keV with a 90\% confidence lower limit of 3.8 keV.
The PSPC temperature would also be lower than those typically
found for clusters of this X-ray luminosity.
These discrepancies and the poor fit to a single temperature
model suggest that the cluster has at least two temperature
components in its spectrum.
Most of the emission (particularly on the large scales sampled
by the MPC) would be in a hard component, while a softer
component would provide much of the flux in the PSPC band.
The residuals in Figure~\ref{fig:whole_spectrum} are also
suggestive of additional soft X-ray emission.
A model with two temperatures does provide an improved fit to the
observations.
The lower temperature is $kT = 1.04^{+0.09}_{-0.15}$ keV, while
the higher temperature is poorly determined but considerably
larger.

Previous $Einstein$ IPC and ROSAT HRI images have shown that
A2597 possesses a cooling flow at the center of the cluster
(Crawford et al.\ 1989; Paper I).
The cooling radius is approximately 100-250 kpc.
Below, we show that the cooler spectral component is localized
to the central regions of the cluster.
Thus, it seems reasonable to attribute the cooler component in
the spectrum of A2597 to this cooling flow, and to model it using
the spectrum of gas cooling subject to its own radiation.
A cooling flow spectrum should be reasonably approximated by
isobaric cooling.
We therefore added a component to the model with the spectrum of
gas cooling isobarically from the ambient cluster gas temperature
to a very low temperature.
The initial temperature and abundances of this cooling flow
component were assumed to be the same as that of the general
intracluster gas.
We refer to this as a ``single temperature plus cooling flow''
model (Table~\ref{tab:spectra}).
This model fit the spectra better than the two temperature
model.

\begin{figure*}[htb]
\vskip4truein
\includegraphics{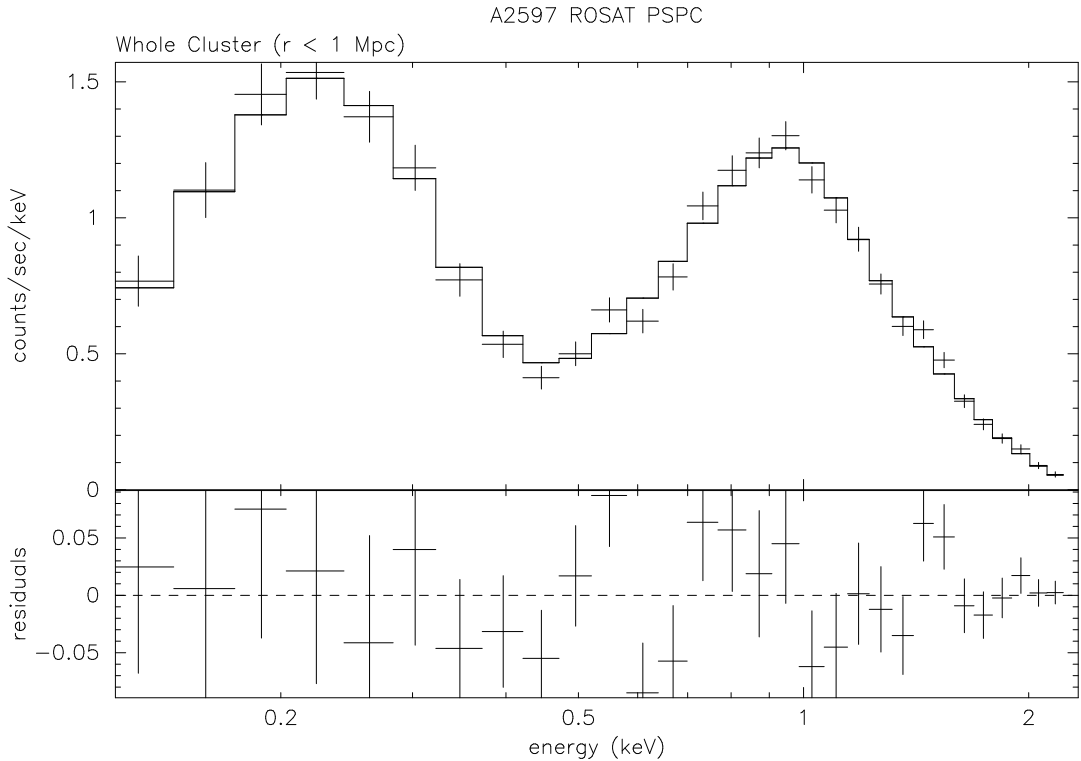}
\caption{The ROSAT PSPC X-ray spectrum for the whole A2597
cluster is compared to the best-fit single temperature plus
cooling flow model.
The notation is the same as in
Figure~\protect\ref{fig:whole_spectrum}.}
\label{fig:whole_cf_spectrum}
\end{figure*}

Figure~\ref{fig:whole_cf_spectrum} shows the best-fit
single temperature plus cooling flow model to the spectrum of
the entire cluster.
This model provides a considerably improved and acceptable fit
to the spectrum of the entire cluster, with
$\chi^2 = 27.05$ for 25 d.o.f.
Some of the parameters of this model are given in
Table~\ref{tab:spectra}.
In this model, the hydrogen column is well-determined, and the
emission measure of the hotter gas and the cooling rate in the
cooling flow are moderately well-determined.
The temperature of the intracluster gas and
its abundances are very poorly determined.
We were only able to derive a lower limit for the gas temperature, and the
abundance is practically undetermined (the errors allow all
sensible values).
This is not surprising for several reasons.
First, the soft X-ray spectrum of isothermal hot gas depends only
weakly on the temperature.
Second, the soft X-ray spectrum of isothermal hot gas is nearly
independent of the abundances because all of the common elements
are fully ionized except iron, and the strong iron lines are at
photon energies of about 7 keV.
Third, the soft X-ray emission from cooling gas is nearly
independent of its initial temperature as long as that
temperature is suitably high
(Wise \& Sarazin 1993).
Fourth, the X-ray spectrum of cooling gas is roughly independent
of the abundances in the gas because increasing the abundances
increases both the emissivity and the cooling rate
(e.g., Wise \& Sarazin 1993).

The cooling rate derived from the spectrum,
$\dot{M} = 344^{+75}_{-67} \, M_\odot$ yr$^{-1}$, is in agreement
with that derived from analyses of the $Einstein$ IPC image
(370 $M_\odot$ yr$^{-1}$; Crawford et al. 1989)
and the ROSAT HRI
(327 $M_\odot$ yr$^{-1}$; Paper I).

Because of the effect of the cooling flow at the center of the
cluster on the overall spectrum,  we determined the
spectrum for the outer intracluster gas in A2597
at projected radii of 250 kpc to 1 Mpc
(Table~\ref{tab:spectra}).
The inner radius of 250 kpc was chosen to be
larger than the cooling radius as determined by
X-ray surface brightness profiles
(Crawford et al.\ 1989; Paper I),
and larger than the region where the spatially resolved spectra,
described below, require a cooling flow component.
The 90\% confidence range for the temperature here
(2.28--6.59 keV) is in better agreement with the
temperature from $Einstein$ MPC measurement.
For the reasons given above, the abundance is poorly determined
at these higher temperatures, and we have only a lower limit of
about 0.43 of cosmic.
The single temperature model provided an adequate fit
to the outer cluster spectra ($\chi^2 = 31.54$ for 26 d.o.f.
The fit was not improved by adding a cooling flow model,
and the upper limit on the cooling rate in the outer cluster
was $\dot{M} < 39 \, M_\odot$ yr$^{-1}$ (90\% confidence).

\begin{figure*}[htb]
\vskip3truein
\includegraphics{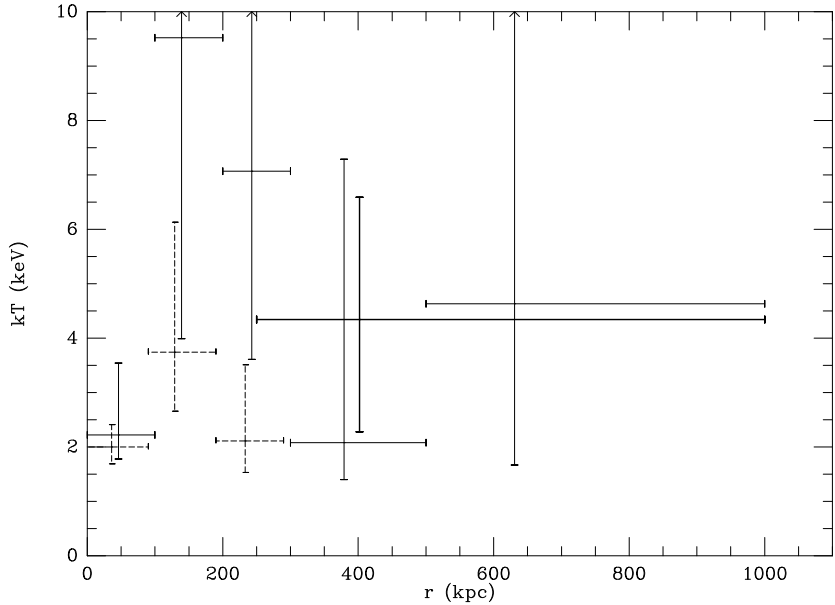}
\caption{The ROSAT PSPC temperature as a function of projected
radius from the X-ray spectra in annuli and the outer cluster spectrum.
The solid error bars are the best models from
Table~\protect\ref{tab:spectra}.
The temperature from the outer cluster spectrum is shown with
a thicker line.
The dashed error bars are the single temperature fits for the
three inner radii where cooling flow components were required
to fit the spectrum.
These points were displaced slightly toward the y-axis to separate
them from the solid error bars;
in fact, they include the same radii.
The vertical error bars are 90\% confidence intervals;
the arrows at the top of the figure indicate temperature values
which are only lower limits.
The horizontal bars give the widths of the regions over which
the spectra were accumulated.
Each point was plotted at the median radius for the X-ray
emission in that annulus.}
\label{fig:temperatures}
\end{figure*}

%
%
\begin{table*}[ht]
\caption[X-Ray Spectral Fits]{}
\label{tab:spectra}
\begin{center}
\begin{tabular}{lcccccccc}
\multicolumn{9}{c}{\sc ROSAT PSPC X-Ray Spectral Fits} \cr
\hline \hline
&\multicolumn{3}{c}{Single Temperature Model}&&
\multicolumn{4}{c}{Single Temperature Plus Cooling Flow} \cr
\cline{2-4} \cline{6-9}
&&&&&&&&\cr
&$N_H$&$kT$&$\chi^2$/d.o.f.&&
$N_H$&$kT$&$\dot{M}$&$\chi^2$/d.o.f.\cr
Region&
($10^{20}$ cm$^{-2}$)&(keV)&&&
($10^{20}$ cm$^{-2}$)&(keV)&($M_\odot$ yr$^{-1}$)&\cr
\hline
Whole Cluster ($r \le 1$ Mpc)&
$1.89^{+0.21}_{-0.20}$&2.61$^{+0.68}_{-0.48}$&47.88/26&&
$1.18^{+0.25}_{-0.15}$&$>$3.52&$344^{+75}_{-67}$&27.05/25\cr
Outer Cluster (0.25--1 Mpc)&
$1.77^{+0.34}_{-0.30}$&$4.34^{+2.25}_{-2.06}$&31.54/26&&
&&&\cr
$r \le 0.1$ Mpc &
$2.49^{+0.14}_{-0.13}$&$2.00^{+0.41}_{-0.31}$&40.88/27&&
$2.48^{+0.14}_{-0.14}$&$2.22^{+1.32}_{-0.44}$&$92^{+85}_{-92}$&38.20/26\cr
0.1--0.2 Mpc&
1.32$^{+0.15}_{-0.16}$&3.74$^{+2.39}_{-1.08}$&29.36/27&&
1.19$^{+0.18}_{-0.12}$&$>$3.99&$81^{+35}_{-35}$&22.67/26\cr
0.2--0.3 Mpc&
1.60$^{+0.24}_{-0.22}$&2.11$^{+1.40}_{-0.58}$&28.58/25&&
1.45$^{+0.24}_{-0.22}$&$>$3.61&58$^{+27}_{-49}$&24.72/24\cr
0.3--0.5 Mpc&
1.95$^{+0.35}_{-0.30}$&2.08$^{+5.21}_{-0.68}$&29.57/27&&
&&&\cr
0.5--1.0 Mpc&
$1.47^{+0.98}_{-0.92}$&$>$1.67&39.40/26&
&&&\cr
\hline
\end{tabular}
\end{center}
\end{table*}

\subsection{Spatially-Resolved Spectra} \label{ssec:spec_annuli}

The PSPC X-ray spectra of A2597 were also determined in a set of
five circular annuli centered on the central cD galaxy
(which is coincident with the peak in the X-ray surface brightness).
The widths were determined from the PSPC
point-spread-function (PSF) in the inner regions, and by requiring
that each spectrum contain at least 1000 net counts in the outer regions.
The resulting fits are shown in Table~\ref{tab:spectra}.
The abundances are not well-determined in any of these spectra,
so we adopted a fixed abundance of 0.5 of cosmic.
This is consistent with the values from the overall spectrum and the
outer cluster spectrum and with the results for other clusters
with similar luminosities
(e.g., Edge \& Stewart 1991).

\begin{figure*}[htb]
\vskip3truein
\includegraphics{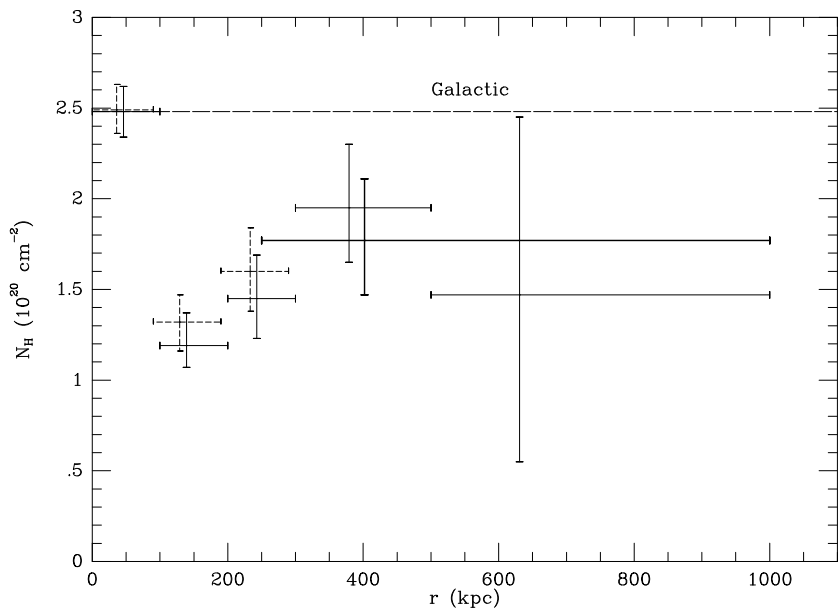}
\caption{The ROSAT PSPC absorbing hydrogen column as a function
of projected radius for the X-ray spectra in annuli.
The notation is the same as in
Figure~\protect\ref{fig:temperatures}.
The dashed horizontal line is the Galactic column from
Stark et al.\ (1992).}
\label{fig:hcol}
\end{figure*}

The spectra in the inner three annuli ($ r \le 300$ kpc)
are not well-fit by single temperature models.
Because the PSPC spectra of these annuli are generally softer
than the $Einstein$ MPC cluster spectrum, we have also fit these
regions including a cooling flow component.
These models are shown at the right of Table~\ref{tab:spectra}.
When the cooling flow components to the spectrum are included,
the ambient gas temperatures in most of the annuli are poorly constrained.
Note that the cooling rate in each of the three inner 100
kpc annuli are similar, suggesting that gas is cooling out
of the hot intracluster medium over a substantial region
of $\sim$250 kpc in radius.
The nearly equal cooling rates in these annuli are crudely
consistent
with $\dot{M} \propto r$, as has been suggested for many other
cooling flow clusters based on surface brightness data
(e.g., Fabian, Nulsen, \& Canizares 1991).
Note that the cooling rate in the central region is quite uncertain.
This is due to the low ambient gas temperature for this region, which
means that both of the two spectral components have similarly soft spectra.
Thus, the cooling flow component makes a smaller contribution
to the ROSAT band, and the ambient gas and cooling flow contributions
are strongly anticorrelated.

For two outer annuli (300-1000 kpc), the fit is not improved by
the addition of a cooling flow component, and the best-fit
cooling rate is zero.
Moreover, the cooling rate for the outer cluster ($r \ge 250$ kpc)
is zero.
Thus, there is no spectral evidence for cooling beyond about 250 kpc.

The temperatures from the fits to these spectra are plotted in
Figure~\ref{fig:temperatures}.
The solid error bars represent either the best single temperature model
or the best single temperature plus cooling flow model for
the three inner annuli where the single temperature model did not give
a good fit.
For the three inner annuli, the single temperature model
is shown as a dashed error bar.
The vertical error bars are 90\% confidence intervals.
The horizontal bars give the width of the annulus over which
the spectrum was accumulated.
Each point was plotted at the median radius for the X-ray
emission in that annulus.
The outer cluster temperature is also shown.
There is no clear evidence for a trend in the ambient gas temperature
except for the drop within 100 kpc.

The hydrogen columns required to fit the soft X-ray absorption
in the PSPC annular spectra are shown in Figure~\ref{fig:hcol}.
The dashed line is the Galactic value from the 21 cm H~I
measurements
of Stark et al. (1992).
Most of the values are somewhat lower than the value from the
radio
observations.
The inclusion of the cooling flow in the inner three spectra doesn't
affect the derived columns in a significant way.

\subsection{Excess Absorption?} \label{ssec:spec_excess}

Excess soft X-ray absorption has been found in the $Einstein$
SSS spectra of many cooling flow clusters
(White et al.\ 1991).
ROSAT PSPC and ASCA spectra have confirmed this absorption
in a number of cases
(Allen et al.\ 1993;
Fabian et al.\ 1994),
and the ROSAT PSPC spectral images have shown that the excess
absorption is confined to the inner cooling regions of the
cluster in some cases
(Allen et al.\ 1993;
Irwin \& Sarazin 1995).
A2597 was not included in the sample studied by White et al.;
however, for strong cooling flows like A2597, the excess columns were
typically $\ga 1 \times 10^{21}$ cm$^{-2}$.

\begin{figure*}[htb]
\vskip3truein
\includegraphics{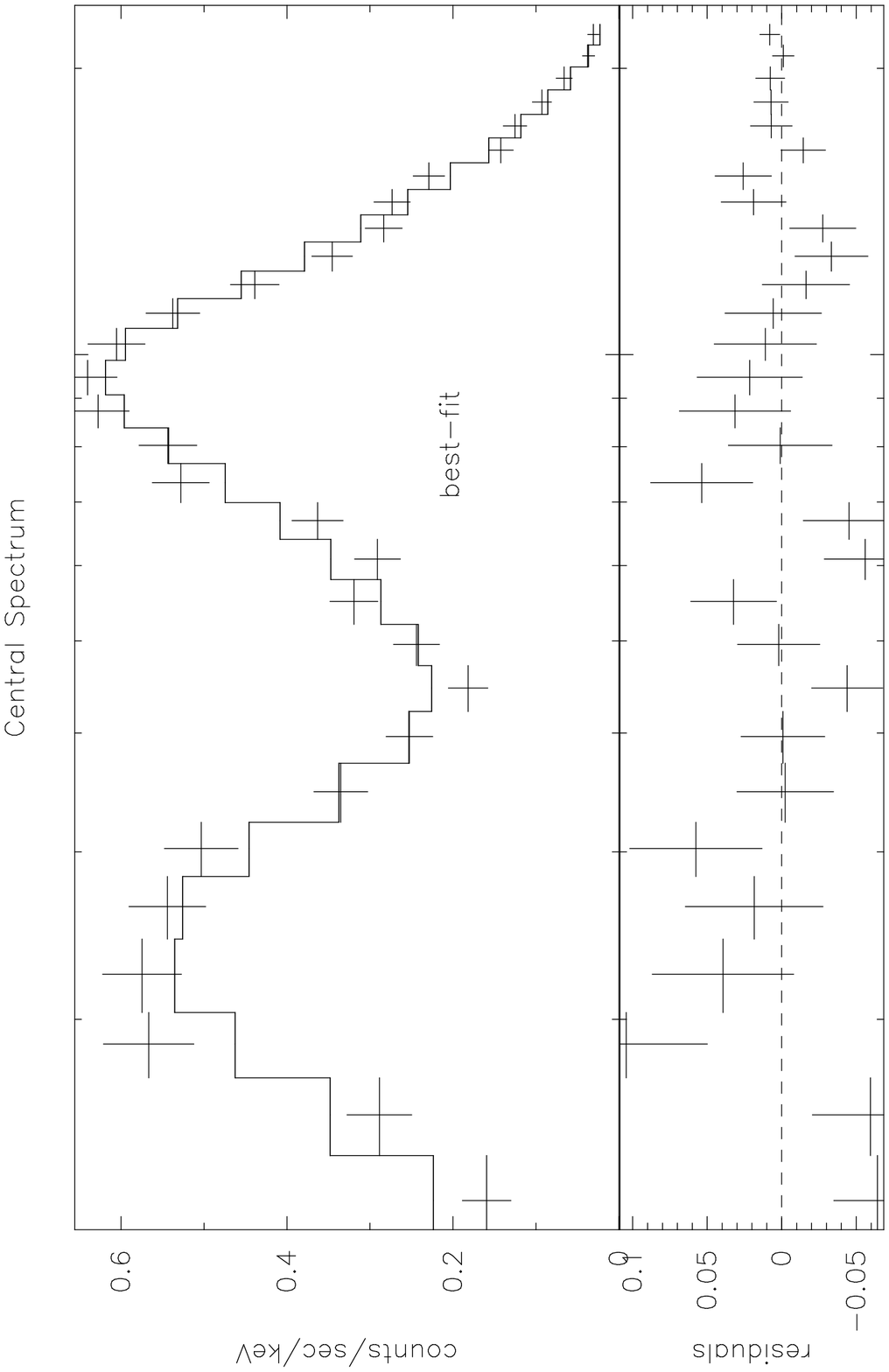}
\vskip3truein
\includegraphics{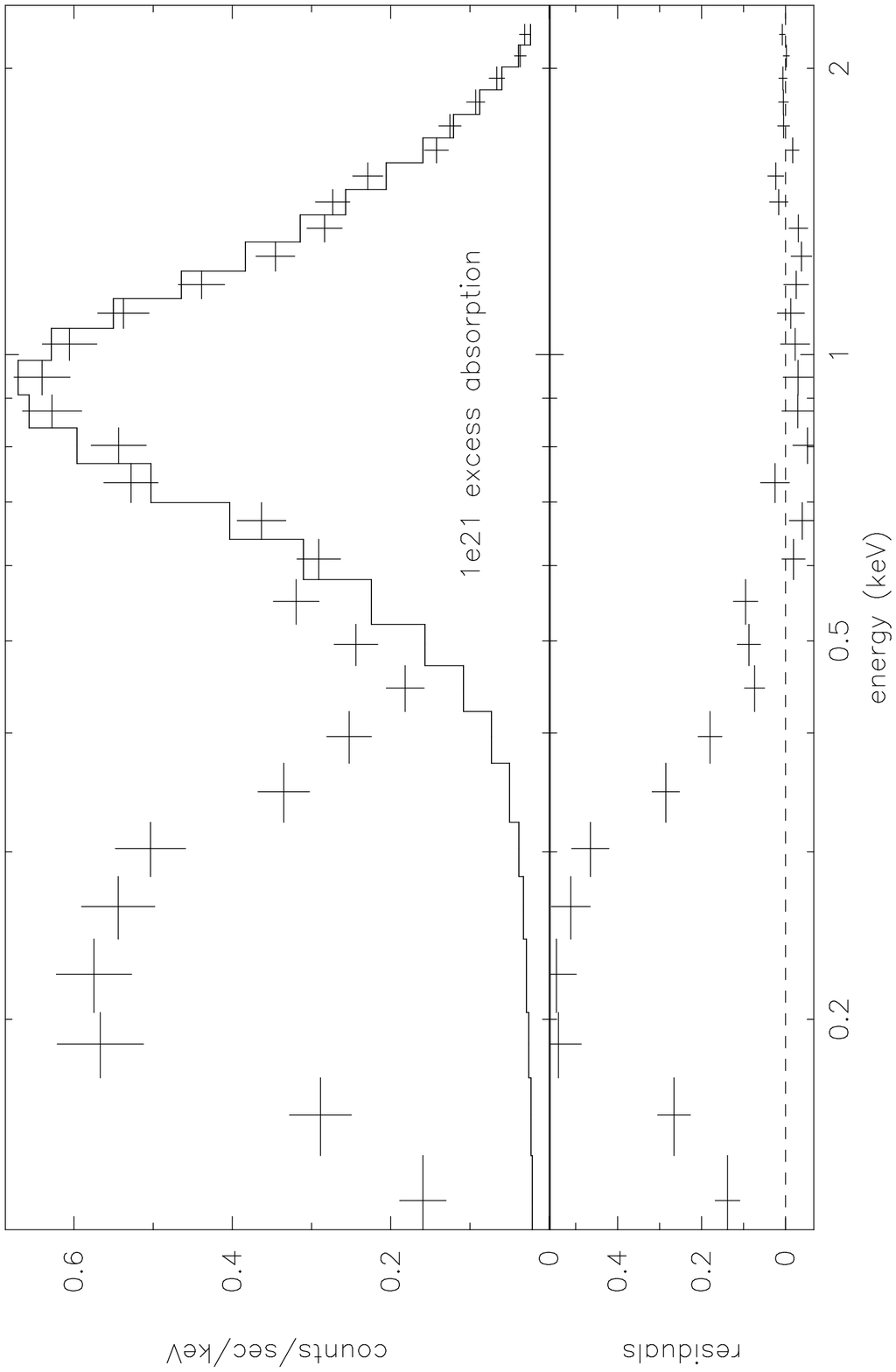}
\caption{(Upper 2 panels):
The ROSAT PSPC X-ray spectrum for the central 100 kpc (in
projected
radius) of the A2597 cluster
is compared to the best-fit single temperature plus cooling flow
model.
The notation is the same as in
Figure~\protect\ref{fig:whole_spectrum}.
(Lower 2 papels):
The same spectrum, but fit with a model with an excess absorption
of $1 \times 10^{21}$ cm$^{-2}$.
}
\label{fig:center_spectrum}
\end{figure*}

Toward A2597, the measured Galactic hydrogen column is
$N_H = 2.45 \times 10^{20}$ cm$^{-2}$
(Stark et al.\ 1992).
There is no evidence for other Galactic interstellar matter from the
IRAS images of the region.
This is a relatively small value, and a significant excess
column should be easy to detect compared with this absorption.
None of the spectra in Table~\ref{tab:spectra} and Figure~\ref{fig:hcol}
require more absorption than the Galactic value, and most require less.
An excess absorption column similar to those seen in other
similar cluster cooling flows ($\Delta N_H \sim 10^{21}$
cm$^{-2}$) can be ruled out with high confidence.
Figures~\ref{fig:center_spectrum}a,b illustrate this point.
Figure~\ref{fig:center_spectrum}a shows the best-fit single
temperature
plus cooling flow model of Table~\ref{tab:spectra} compared to
data
for the inner 100 kpc of the cluster.
Figure~\ref{fig:center_spectrum}b compares this to the best-fit
model
of the same type, but with an excess column of
$\Delta N_H = 10^{21}$ cm$^{-2}$.
The ROSAT PSPC response naturally divides the spectrum into
soft and hard bands at about 0.4 keV.
In the presence of excess absorption on the order of
$\Delta N_H \sim 10^{21}$ cm$^{-2}$, the soft band should be
strongly suppressed.
This is not observed in the central spectrum of A2597
(Fig.~\ref{fig:center_spectrum}a).
The best-fit model with an excess column of
$\Delta N_H = 10^{21}$ cm$^{-2}$
has an increase in $\chi^2 = 662$ for one less degree of freedom,
and is thus completely excluded.

\begin{figure*}[htb]
\vskip3truein
\includegraphics{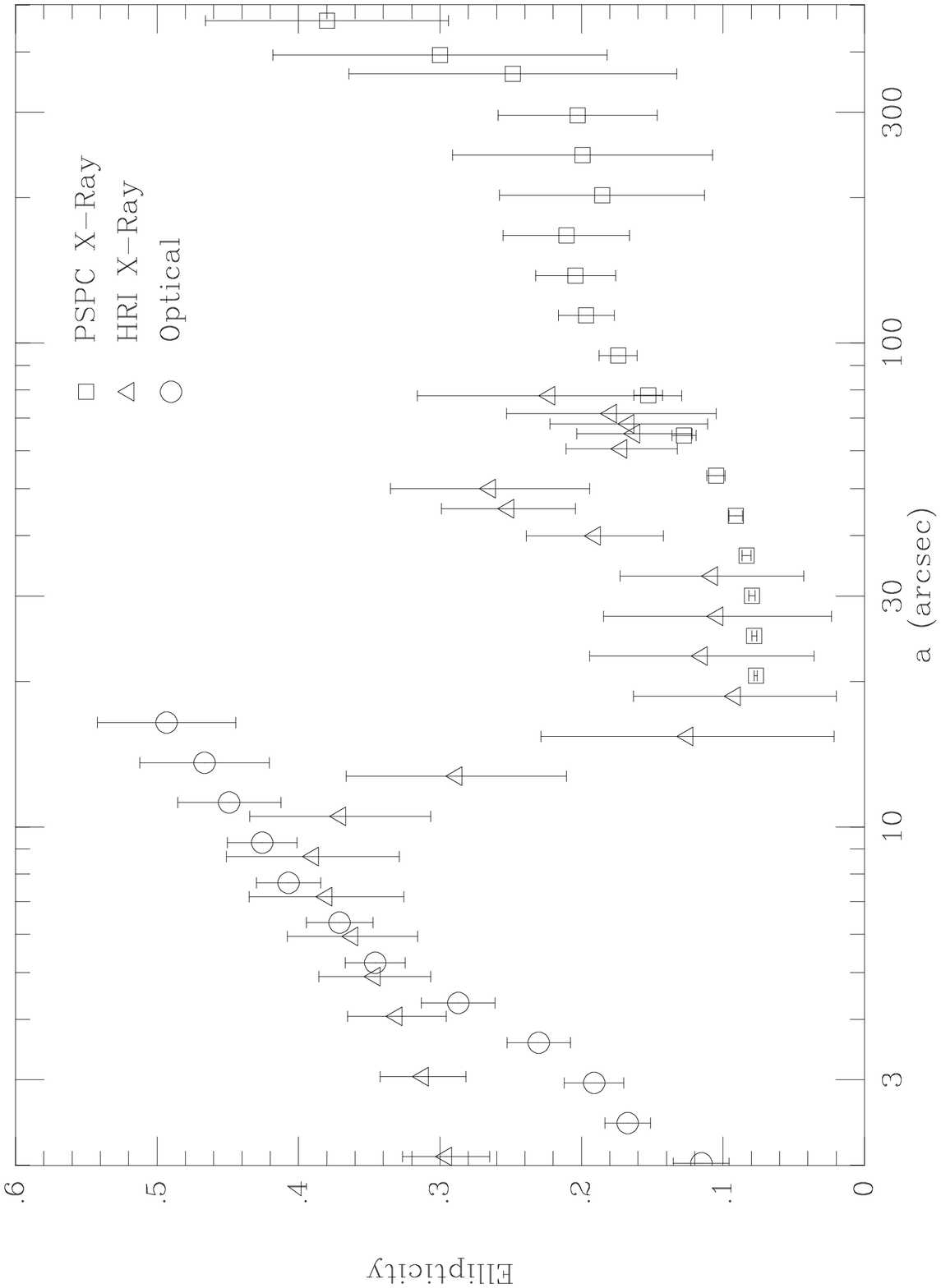}
\caption{The ellipticity of the X-ray and optical isophotes of
A2597 are shown as a function of the semimajor axis $a$
of the isophotes.
The squares and triangles are from ROSAT PSPC and HRI images,
respectively.
The HRI data is from Paper I.
The circles are from the optical V-band image of the central
cD galaxy from McNamara \& O'Connell (1993).
One $\sigma$ error bars are shown.}
\label{fig:eps}
\end{figure*}

The central spectrum in Figure~\ref{fig:center_spectrum} was
collected from within a projected radius of 100 kpc.
Thus, it includes emission from regions at larger radii in front
of and behind the cluster core.
If there were excess absorption associated with the gas in the
core, then it would act only on the emission within and behind the
core.
This might dilute the effect of the excess absorption.
To try to measure or limit the amount of absorption acting
on the emission from the cooling core alone, we constructed
a spectrum in which the background was taken from the outer
annuli, and scaled so as to remove the foreground cluster emission.
The scaling was determined from the surface brightness profile
of the cluster (\S~\ref{ssec:surf_bright}).
This did produce a somewhat better single temperature fit to the
spectrum ($\chi^2 = 36.75$ for 27 d.o.f.) with a lower temperature
($kT = 1.75^{+0.46}_{-0.21}$) and a slightly higher absorbing
column ($N_H = 2.79^{+0.22}_{-0.20}$).
The column was identical in a model with a cooling flow.
Since this gave the highest value of the central absorbing column
of any of the methods employed, we use this value to set an upper
limit on the excess absorption associated with the cluster cooling
core.
If this value is compared to the measured Galactic column, the
difference is $0.54 \times 10^{20}$ cm$^{-2}$.

There may be systematic errors in converting the wide-beam
H~I measurements of
Stark et al.\ (1992)
to a total hydrogen column toward the cooling core of A2597.
To give more conservative upper limit to the excess, we compare
the absorbing column for the single temperature model in front of
the cooling core with the value for outer cluster region
(Table~\ref{tab:spectra}).
Note that this outer spectrum should be representative of the
foreground absorption in the outer regions of the cluster.
The abundances were again fixed at 50\% of cosmic.
Including the effects of the redshft, the
90\% confidence upper limit on the excess absorption column was
\begin{equation} \label{eq:dNH}
\Delta N_H < 1.68 \times 10^{20} \, {\rm{cm}}^{-2} \, .
\end{equation}
Given the assumptions that were made to maximize the value of the
excess absorption, this would seem to be an extreme upper limit.
Thus, the excess absorption toward A2597 is at least an order of
magnitude smaller than typical values toward other similarly
large cooling flows
(White et al.\ 1991).

Remarkably, A2597 is one of the few cooling flows toward which
a large column has been detected in 21 cm H~I absorption
(O'Dea et al.\ 1994a).
O'Dea et al.\ detect two absorption components.
A narrow line is seen toward the nucleus
with a velocity width of $\approx$220 km s$^{-1}$ and a column of
$N_H^n = 8.2 \times 10^{20} ( T_s^n / 100 \, {\rm{K}})$
cm$^{-2}$, where $T_s$ is the spin temperature of the hydrogen.
This component is not spatially resolved.
There is also a broad component with a width of $\approx$410 km
s$^{-1}$ which appears extended over the entire radio source.
This broad component has a column of
$N_H^b = 4.5 \times 10^{20} ( T_s^b / 100 \, {\rm{K}})$
cm$^{-2}$.
O'Dea et al.\ suggest that the broad component is associated with
the optical emission line nebula at the center of the the cooling
flow.

\begin{figure*}[htb]
\vskip3truein
\includegraphics{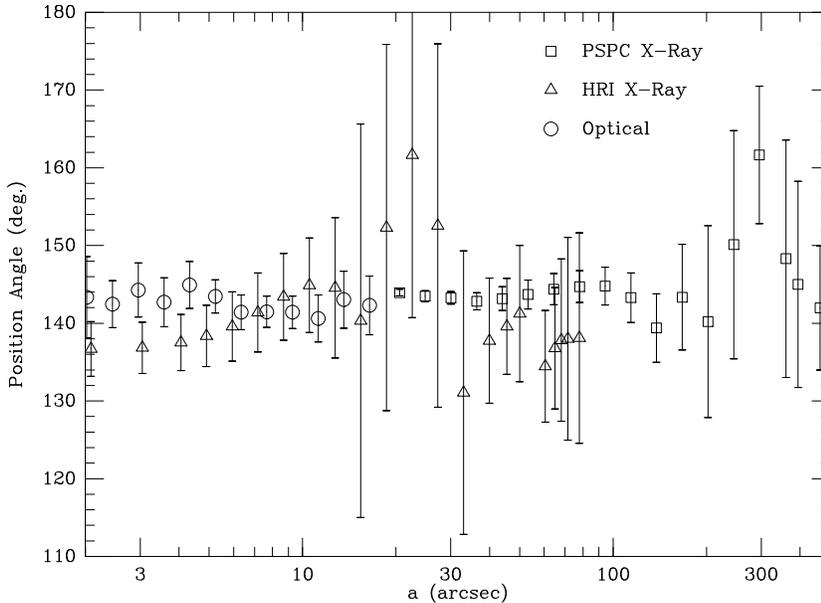}
\caption{The position angles of the X-ray and optical isophotes
of
cluster A2597 are shown as a function of the semimajor axis $a$
of the isophotes.
The position angle is measure from north toward the east.
The notation is the same as in Figure~\protect\ref{fig:eps}.}
\label{fig:pa}
\end{figure*}

Is our upper limit on the X-ray absorption consistent with the
detected hydrogen in front of the radio source in this cD galaxy?
The radio source is extremely small, extending only to a
projected radius of $r < 4$ arcsec.
The absorption could be greatly diluted over the larger aperture
used for the PSPC central spectrum.
The PSF of the PSPC does not allow one to determine the spectrum
on the scale of the radio source.
However, to test the consistency between the X-ray spectrum and
the radio data, we compared the X-ray data to a spectral model
in which the total column,
$N_H = 12.7 \times 10^{20}$ cm$^{-2}$, associated with the
H~I absorption at a spin temperature of $T_s = 100$ K covers the
inner 4 arcsec of the X-ray emission.
This absorption was in addition to the previously
determined absorption (Table~\ref{tab:spectra}).
This exaggerates the effect of the radio absorption in several
ways.
First, the narrow 21 cm component is spatially unresolved, and
covers a much smaller area.
Second, part of the radio absorption was presumably included in
the previous spectral fit.
The fraction of the X-ray emission from this area is estimated
from the surface brightness profile of the ROSAT HRI
(\S~\ref{ssec:surf_bright}; Paper I), corrected for
the PSF of the HRI.
All other spectral parameters were kept fixed at their previous
values (Table~\ref{tab:spectra}).
This model spectrum could not be distinguished from the previous
best-fit model by even 1-$\sigma$.
Thus, our X-ray spectra and the upper limit on excess absorption
we derived are consistent with the H~I absorption detections of
O'Dea et al.\ (1994a).

O'Dea et al.\ suggest that the broad absorption component is
associated with the optical emission line nebula, which covers
a much larger projected area of $r < 14$ arcsec
(Heckman et al.\ 1989).
For a spin temperature of 100 K, the broad component has a column
density of $N_H^b = 4.5 \times 10^{20}$ cm$^{-2}$.
Such a column of hydrogen covering the inner 14 arcsec 
radius of the X-ray emission might be detectable.
Simply adding this absorption to the model in
Table~\ref{tab:spectra} increases the $\chi^2$ by 27.
If one refits the spectrum including this absorption by the broad
H~I component in the inner 14 arcsec, the foreground column is
reduced from 2.49 to $2.07 \times 10^{20}$ cm$^{-2}$.
The spectral fit is worse by $\Delta \chi^2 = 2.95$, which can be
rejected with more than 90\% confidence.
Thus, the X-ray spectra are moderately inconsistent with the
broad component having a column density of
$N_H^b = 4.5 \times 10^{20}$ cm$^{-2}$ that uniformly covers the
entire region of the emission line nebula ($ r < 14$ arcsec).
A larger region of coverage is certainly ruled out by the
upper limit of $2.62 \times 10^{20}$ cm$^{-2}$ on the column
in the central X-ray spectrum.
On the other hand, the spin temperature may be less than 100 K,
the abundance of carbon and oxygen (which provide a large part
of the X-ray absorption) may be lower than cosmic, or
the H~I may be centrally condensed (as the H$\alpha$ emission
is observed to be) and the neutral column from the radio
observation may only apply at the center.
It is likely that a centrally condensed distribution of neutral
hydrogen similar to the surface brightness of the H$\alpha$
could not be ruled out by our PSPC spectrum.

\section{SPATIAL ANALYSIS} \label{sec:spatial}

\subsection{Isophotal Shape} \label{ssec:ellipse}

Figure~\ref{fig:xray_cluster} shows that the X-ray emission from
A2597 is elongated.
We measured the shape of the X-ray image using the best-fit elliptical
isophotal model for the X-ray emission.
The X-ray surface brightness was assumed to be represented by
a series of concentric elliptical isophotes.
The centers of the isophotes were fixed at the X-ray centroid,
while the ellipticities and position angles
of the fitted ellipses were allowed to vary.
The elliptical isophotes were determined using the algorithm of
Jedrzejewski (1987).
A similar analysis was done previously for the ROSAT HRI image
of A2597 (Paper I).
For comparison, we determined the ellipticity
of the central cD galaxy in A2597 using the V-band image from
McNamara \& O'Connell (1993) in a similar manner.

Figure~\ref{fig:eps} shows the ellipticity, $\epsilon$, as a
function of the the semimajor axis, $a$, of the elliptical
isophotes for the ROSAT PSPC, ROSAT HRI, and optical V-band
data.
At the smallest semimajor axes plotted for each instrument
($a \la 2^{\prime\prime}$ in the optical,
$a \la 4^{\prime\prime}$ in the
HRI, and $a \la 30^{\prime\prime}$ in the PSPC),
the ellipticity may be reduced somewhat by the instrument PSF.
The ROSAT HRI image is strongly elongated ($\epsilon \approx 0.4$)
at the center;
this was discussed in some detail in Paper I.
At larger radii ($a \ga 15$ arcsec), the X-ray ellipticity is much
smaller ($\epsilon \approx 0.1$), although it increases to larger
radii.
The ROSAT PSPC and HRI elongations are consistent where they
overlap, except for a narrow range of radii around 45 arcsec.
This might be due to a weak, variable X-ray source.
The PSPC image becomes fairly elongated at the largest radii for
which the shape could be measured with any confidence.

Figure~\ref{fig:pa} gives the position angle of the
semimajor axis (measured north through east)
for the three images.
We see that the orientation of the X-ray isophotal ellipses
remains nearly constant from 2 arcsec out to 500 arcsec.
The position angle may increase slightly with radius
(from 137$^\circ$ to approximately 145$^\circ$).
The ellipticities of the V-band isophotes of the cD galaxy increase
rapidly with radius in the inner 7 arcsec;
the ellipticity is nearly constant at values between 0.4-0.5 at
larger radii (Figure~\ref{fig:eps}).
Increasing ellipticity with radius is a common property of brightest cluster
ellipticals
(Porter, Schneider, \& Hoessel 1991).
The isophotal position angles are fairly constant with radius
at a value of $\approx$$143^\circ$.
The isophotal ellipticities and position angles found from our V and
I data are concordant,
as are those in the Thuan-Gunn R-band by
Porter et al.\ (1991).
We have therefore presented only the V-band data here.
Although some of the increasing ellipticity with
radius seen in the optical band is probably real, the slope and
intercept (central ellipticity and position angle) may
be strongly affected by seeing in the inner several arcsec.
Franx, Illingworth, \& Heckman (1989)
showed that optical ellipticities and position angles can be
strongly affected  (greater than 10\%) by the width and shape of the
instrument PSF within about 5 times the seeing $FWHM$.
The seeing for the V and I band data was $FWHM = 1.6$ arcsec.
The position angles of the optical and X-ray data agree to
within their uncertainties, and the average optical
and X-ray ellipticities within 15 arcsec are similar.

On large scales, the orientation of the galaxy distribution in
A2597 is similar to that of the X-ray emission
(Tr\`evese, Cirimele, \& Flin 1992a,b).
The ellipticity of the galaxy distribution is somewhat larger than
that of the X-ray emission, although the errors on both are also
quite significant.
Buote \& Canizares (1996) note that the ellipticity of the mass
distribution and galaxy distribution in clusters is expected to
exceed that of the X-ray distribution in most cases.

\begin{figure*}[htb]
\vskip3truein
\includegraphics{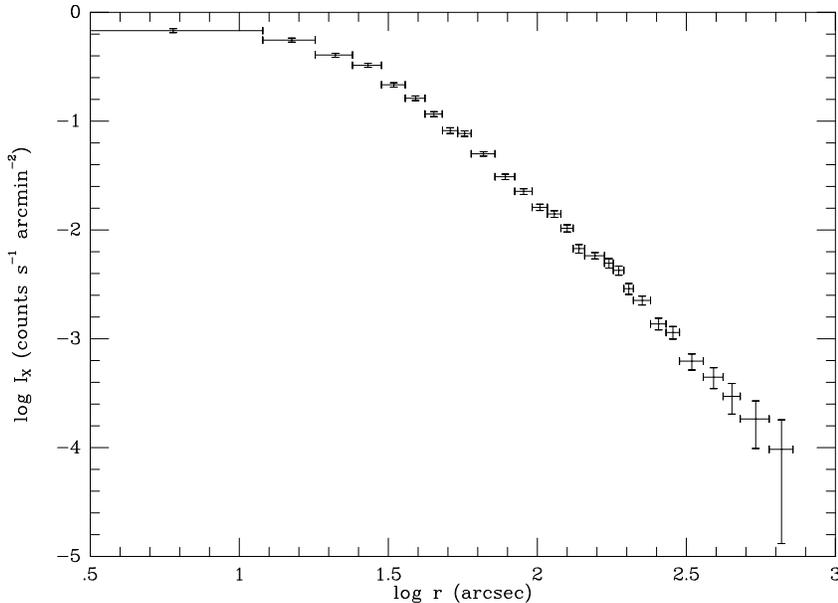}
\caption{The ROSAT PSPC X-ray surface brightness profile of
A2597, corrected for background and exposure.
This surface brightness in counts s$^{-1}$ arcmin$^{-2}$ in the
photon energy band of 0.4 -- 2.4 keV (PI channels 42--247).}
\label{fig:xprofile}
\end{figure*}

In order to search for substructure within the ROSAT PSPC image of
A2597, we constructed a synthetic image of the cluster using the data
on the best-fit elliptical isophotes of the cluster in
Figures~\ref{fig:eps} and \ref{fig:pa}.
This was subtracted from the actual image of the cluster in
Figure~\ref{fig:xray_cluster}, and the significance of any
residuals was determined.
(This procedure is described in more detail in Paper I.)
The only significant positive features were associated with
Srcs.~9, 19, 20, 21, and 23 in Table~\ref{tab:sources}, and with
two weaker (4 $\sigma$) sources at
R.A.\ = 23$^{\rm h}$24$^{\rm m}$43{\fs}7 and
Dec.\ = $-$12{\arcdeg}07{\arcmin}31{\arcsec}, and
R.A.\ = 23$^{\rm h}$25$^{\rm m}$56{\fs}2 and
Dec.\ = $-$12{\arcdeg}09{\arcmin}24{\arcsec} which are seen at
the center right and left of Figure~\ref{fig:xray_cluster}.
There are a few weak, negative features at the northern and
southern ends of the cluster, but other than this,
the cluster is very well fit by elliptical isophotes.
This suggests that the cluster is well-relaxed and in hydrostatic
equilibrium out to projected radii of 1---2 Mpc.
In the HRI image of the very center of the cluster cooling flow,
significant substructures are seen (Paper I).
There is an elongated bar of emission which appears partially
as the central increase in the ellipticity in
Figures~\ref{fig:eps}.
There are also some other asymmetries in the emission near the
center.
These are apparently unresolved in the PSPC image.

\subsection{X-Ray Surface Brightness Profile}
\label{ssec:surf_bright}

The azimuthally averaged X-ray surface brightness distribution of
the cluster was derived from the ROSAT PSPC counts in circular
annuli centered on the X-ray centroid.
The photon energy band of 0.4 -- 2.4 keV (PI channels 42--247)
was used to minimize the background contribution
(which is mainly very soft) compared
to cluster emission (which is expected to be mainly very hard).
Although the X-ray image of the cluster is moderately elliptical
(\S~\ref{ssec:ellipse}),
the use of elliptical annuli with variable ellipticities and/or
position angles would make it difficult to deconvolve the
gas density from the X-ray surface brightness.
In addition to quantities derivable from the observed surface
brightness, one would need to specify the third axis of the ellipsoid, and
two additional angles.
Instead, we use circular annuli.
White et al.\ (1994) have shown that small ellipticities of the
order of those in A2597 do not strongly affect the inferred X-ray
surface brightness and gas density.
In general, the surface brightness of an elliptical isophote
corresponds approximately to the azimuthally-averaged surface
brightness at a radius equal to the geometric mean of the
semimajor and semiminor axes of the ellipse.

The resulting ROSAT PSPC surface brightness profile, corrected
for background and exposure, is shown in Figure~\ref{fig:xprofile}.
Point sources have been removed.
X-ray emission is detected out to a radius of
$1080'' = 18' = 2.37$ Mpc.
However, the errors in the surface brightness are very large beyond
$12' = 1.58$ Mpc, and we only include data within this radius in
Figure~\ref{fig:xprofile}.

\begin{figure*}[htb]
\vskip4truein
\includegraphics{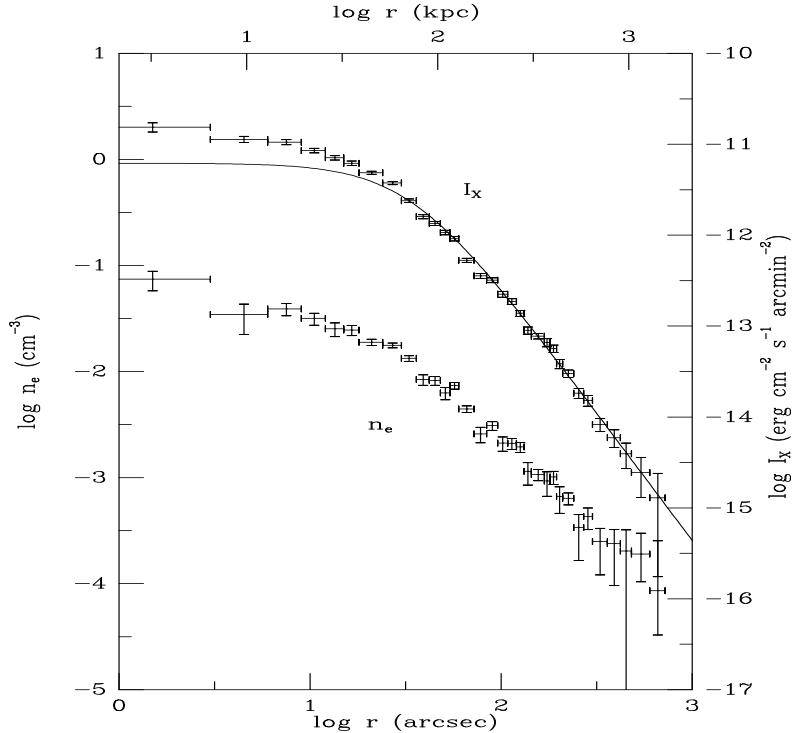}
\caption{The merged ROSAT PSPC and HRI X-ray surface brightness
profile and de-projected electron density for A2597.
The upper data points give the azimuthally averaged X-ray surface
brightness in the band of 0.4 -- 2.4 keV as incident at the
Earth, corrected for absorption.
The scale $I_X$ is given on the right--hand axis.
The best beta model fit, excluding data in the inner 84 arcsec,
is shown as the solid curve.
The lower data points give the de-projected electron number
density;
the scale is given on the left--hand axis.
For each curve, the vertical error bars give the 1 $\sigma$
errors.
Note that the deconvolution causes the errors in the electron
density values to be correlated.
The radial error bars show the widths of the annuli.
The lower axis gives the radial scale in seconds of arc, while
the upper axis gives the scale in kpc.}
\label{fig:x-ray_surf}
\end{figure*}

At smaller radii, the ROSAT HRI spatial resolution is better.
Because of the longer exposure in the HRI, the counts in the
central regions of the image are similar, and so the errors in
the surface brightness there are similar.
At large radii, the lower background in the PSPC
provides a better determination of the X-ray surface brightness.
At intermediate radii (1--4 arcmin), the HRI and PSPC give
consistent results.
To take advantage of the better resolution of the HRI at small
radii and the lower background of the PSPC at larger radii, we
have merged the two profiles into a single X-ray surface
brightness profile.
We adopt the HRI profile for projected radii $r < 108$ arcsec
and the PSPC profile for projected radii $r > 108$ arcsec.
(The HRI profile was published previously in Paper I.)
To match the two, we converted the count rate in the HRI to the
equivalent value in the PSPC, we adopted the region from 200--300
kpc and determined the conversion from the instrument responses,
assuming the best-fit single temperature spectrum for this region
from Table~\ref{tab:spectra}.
The counting rate was converted into a physical flux assuming the
best-fit single temperature spectrum for the entire cluster
(Table~\ref{tab:spectra}).
However, the adopted spectra have little actual affect on the
scaling of HRI to PSPC counts or the conversion to physical flux.
The X-ray surface brightness profile is shown as the upper points
in Figure~\ref{fig:x-ray_surf} with 1 $\sigma$ errors.

We fitted the X-ray surface brightness in
Figure~\ref{fig:x-ray_surf}
with the isothermal ``beta'' model,
$I_X (r) = I_o [ 1 + ( r / r_{core} )^2 ]^{-3 \beta + 1/2}$.
When all of the data points in Figure~\ref{fig:x-ray_surf} were
included, the best-fit values were $r_{core} = 21 \pm 2$ arcsec and
$\beta = 0.59 \pm 0.09$, but
the fit was not acceptable
(minimum $\chi^2 = 59.99$ for 29 d.o.f.).
As had generally been found with cooling flows clusters
(e.g., Jones \& Forman 1984),
the beta model could not fit the sharply peaked central surface
brightness within the cooling radius.
We tried a number of simple models for the cooling flow portion
of the X-ray surface brightness profile, but none gave an acceptable
fit.
Instead, we fit the surface brightness with the beta model, progressively
removing the interior points until an acceptable fit was found.
This required the removal of the data points
interior to a projected radius of 84 arcsec (180 kpc).
This is similar to the cooling radius in the cluster
(Crawford et al.\ 1989; Paper I).
With these interior points removed, an acceptable beta model fit
was found ($\chi^2 = 12.88$ for 14 d.o.f.), which gave
$\beta = 0.64^{+0.08}_{-0.03}$ (90\% confidence).
Note that this value of $\beta$ is consistent, within the errors,
with that found from the unacceptable fit including all of the data
points.
This is not surprising, as the value of $\beta$ is mainly constrained
by the decline in the X-ray surface brightness at large radii, and
the data points at large radii were included in both fits.
The fit excluding the central points gave only an upper limit to the
core radius, $r_{core} < 78$ arcsec (90\% confidence).
Again, this upper limit on the core radius includes the value
found in the unacceptable fit including all of the data points.
Thus, it seems fair to conclude that the centrally peaked X-ray emission
due to cooling flow makes it impossible to determine the core radius of
the cluster, and that the data only allow an upper limit of
$r_{core} < 78$ arcsec (170 kpc).
On the other hand, the value of $\beta$ is well-determined, and the
behavior of the surface brightness at large radii is consistent with the
beta model.
The best-fit beta model, excluding the central points, is shown
as a solid curve in Figure~\ref{fig:x-ray_surf}.
A de-projection of the surface brightness of the cooling flow region
was given previously in Paper I.

Based on numerical simulations of clusters,
Navarro, Frenk, \& White (1995) and Bartelmann \& Steinmetz (1996) suggest
that the actual density profiles of the intracluster gas steepen as
compared to the beta model fits.
Figure~\ref{fig:x-ray_surf} doesn't show any evidence for this, but
our data may not extend to a large enough radius.

The X-ray emissivity of the gas was determined by de-projecting
the X-ray surface brightness into spherical shells, each of
constant emissivity
(e.g., Arnaud 1988).
The electron density was then determined and is plotted
as the lower points in Figure~\ref{fig:x-ray_surf}.

\section{GAS AND TOTAL MASS PROFILES} \label{sec:masses}

\begin{figure*}[htb]
\vskip4truein
\includegraphics{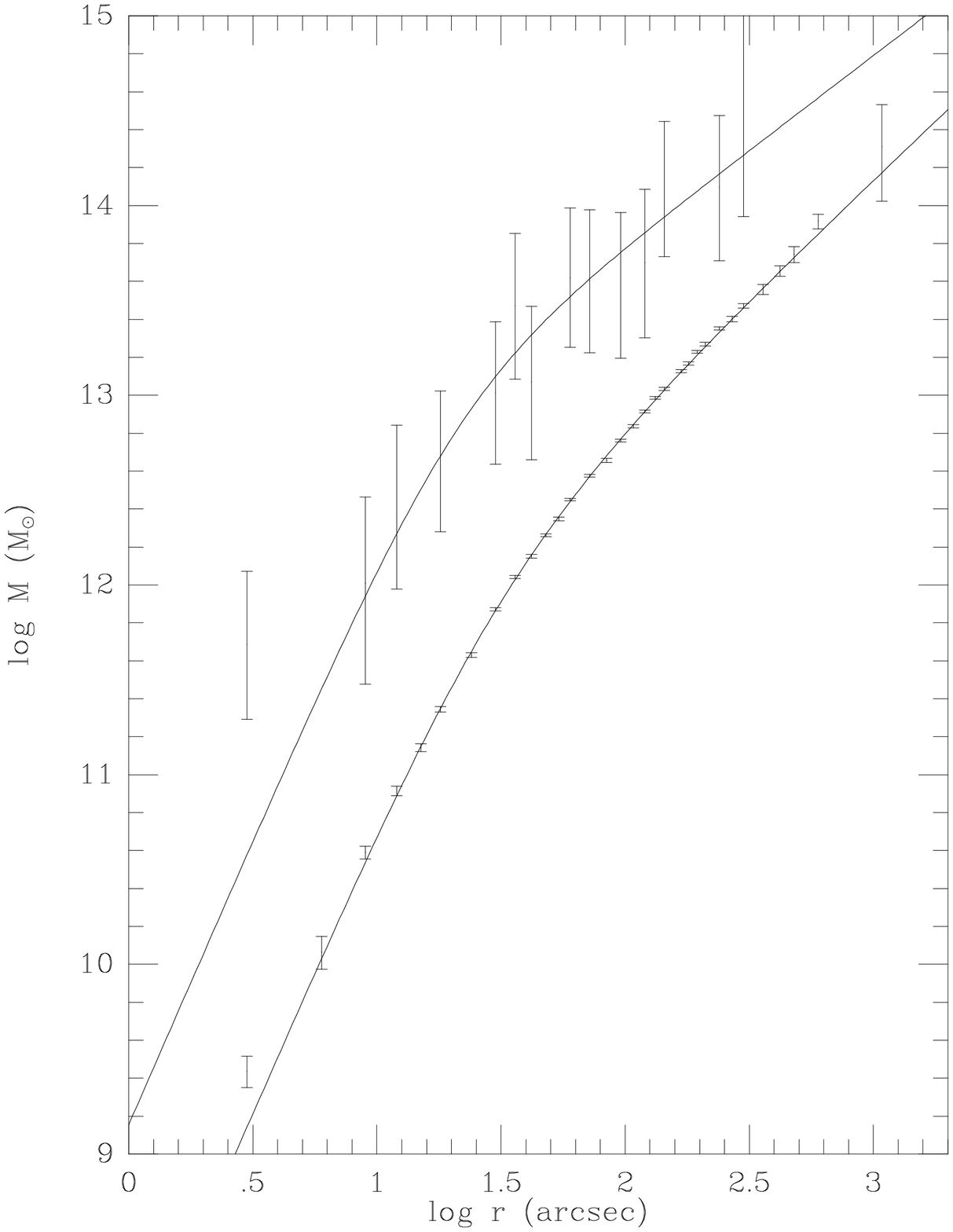}
\caption{The profile of the gas masses (lower) and total gravitational
masses (upper) as derived from the ROSAT merged PSPC and HRI X-ray surface
brightness profile
Fig.~\protect\ref{fig:x-ray_surf}.
The data points give the masses derived from the deconvolved gas density
points in Fig.~\protect\ref{fig:x-ray_surf}.
For the total mass (upper points), some of gas density points were
combined into larger bins to give a monotonically decreasing pressure
profile.
Note that the deconvolution causes the error in the masses to be correlated
for both the gas mass and total mass.
The solid curves are the masses found assuming the best-fit beta-model
X-ray surface brightness profile for the cluster
(\S~\protect\ref{ssec:surf_bright}).}
\label{fig:masses}
\end{figure*}

The values of the electron density from
Figure~\ref{fig:x-ray_surf} were integrated over the volume
to give the profile of the gas mass interior to the
radius $M_{gas} (r)$ (Figure~\ref{fig:masses}).
The total gravitational mass was determined from the
assumption that the gas is in hydrostatic equilibrium.
This implies that the total gas mass interior to each
radius is given by
\begin{equation}
M_{tot} (r) = - \frac{r^2}{G \rho_{gas}} \, \frac{d P}{d r}
\, ,
\label{eq:hydrostatic}
\end{equation}
where $P$ and $\rho_{gas}$ are the gas pressure and density,
respectively.
The gas density $\rho_{gas}$ was taken from
Figure~\ref{fig:x-ray_surf}.
To determine the gas pressure, we need to know the gas
temperature as a function of radius in the cluster.
The gas temperature as a function projected radius is
shown in Figure~\ref{fig:temperatures} as derived from
the PSPC spectra.
It is clear that the errors are very large, and that
the temperature is very poorly constrained in the
inner points where there is a contribution to the
spectrum from a cooling flow.
There is no clear evidence for a radial trend in the temperatures,
except for the drop in temperature within the inner 100 kpc.
Thus, we have assumed a constant value for the temperature set by
the outer cluster spectrum,
$k T = 4.34^{+2.25}_{-2.06}$ keV.
The errors in the masses were assessed using
Monte Carlo simulations of the X-ray surface brightness data to determine
the 90\% confidence region, including
the errors on the temperature in each data point individually.
Note, however, that the errors for the points are correlated in a complex
manner, both because of the de-projection of the surface brightness and
of the inclusion of the error in the overall cluster temperature separately
in the error in the mass at each radius.
Obviously, ASCA spectra of A2597 would be of great value in determining
the temperature profile more accurately.

For comparison, we also determined the gas and total masses
for the best-fit beta-model surface brightness fits for the
gas distribution.
We used the best-fit model using all of the
surface brightness points, including the central excess emission
in the cooling flow.
The resulting mass profiles for the best-fit model are
the solid curves in Figure~\ref{fig:masses}.
The de-projection densities are not corrected for the effects of the
finite instrument PSF, and this causes the de-projection to give different
results than the best-fit beta-model (which is corrected for the PSF) for
the innermost points.

The mass profiles for the best-fit beta-model give
$M_{gas} = 1.2 \times 10^{14}$ $M_\odot$ and 
$M_{tot} = 5.6 \times 10^{14}$ $M_\odot$ at  $r = 2$ Mpc,
for a gas mass fraction of about 21\%.
Similarly, the values at a radius of  $r = 1$ Mpc are
$M_{gas} = 5.0 \times 10^{13}$ $M_\odot$ and
$M_{tot} = 2.8 \times 10^{14}$ $M_\odot$, so that
$M_{gas} / M_{tot} = 18$\%.
The uncertainty in the spectrally-determined temperatures 
mean that the values of the total
mass are quite uncertain (by at least a factor of two).
The gas masses are more accurately known.
For example, if we assume that the gas is isothermal
at a temperature of 9.1 keV (David et al.\ 1993), then
the gas mass would be increased by about 6\%, while the
total mass would increase by a factor of 2.1.

\section{CONCLUSIONS} \label{sec:conclusion}

We have presented an analysis of the ROSAT PSPC and HRI
X-ray observations of the cooling flow cluster A2597.
The X-ray emission from the cluster extends to at least 2 Mpc.
The X-ray image of A2597 is moderately elliptical, and
is elongated in the same direction as optical image of the
central cD galaxy and the galaxy distribution in the cluster.
The ROSAT PSPC X-ray image is well-represented by regular elliptical 
isophotes, suggesting that the cluster is reasonably relaxed and in
hydrostatic equilibrium.

We analyzed the ROSAT PSPC X-ray spectra of the cluster.
Both the overall cluster spectrum and the spatially resolved spectra
within 300 kpc require the presence of both hot and cool components.
The cool component is fit reasonably by a cooling flow spectral model.
The spectrally determined total cooling rate of
$\dot{M} = 344^{+75}_{-67} \, M_\odot$ yr$^{-1}$ is in good
agreement with those derived from analyses of the $Einstein$ IPC
(370 $M_\odot$ yr$^{-1}$; Crawford et al. 1989)
and the ROSAT HRI
(327 $M_\odot$ yr$^{-1}$; Paper I)
X-ray surface brightness profiles.
The spatially resolved X-ray spectra indicate that the cooling component
is distributed over the inner 300 kpc in radius of the cluster,
with significant amounts of gas cooling in the outer parts of this range.
This is in reasonably good agreement with the results of analyses of
the X-ray surface brightness distribution of the cluster
(Crawford et al.\ 1989; Paper I),
and with the general result that gas cools in an extended region
in other cooling flow clusters
(e.g., Fabian, Nulsen, \& Canizares 1991).

If the cooling flow component is not included in the fits to the
ROSAT PSPC spectra, the fits are generally poor and the gas temperatures
are too low to be consistent with the $Einstein$ MPC spectrum of the
entire cluster
(David et al.\ 1993).
When the cooling flow component is included, the gas temperature is poorly
constrained but generally consistent with the MPC value.
As a result, the ROSAT PSPC spectra do not lead to an accurate profile
of the variation of the ambient cluster gas temperature with radius.
Hopefully, ASCA spectra will provide this information.

We do not detect any significant excess absorption toward the center
of A2597.
We find a very conservative upper limit
of the excess absorbing column at the cluster cooling flow to be
$\Delta N_H < 1.72 \times 10^{20}$ cm$^{-2}$.

Remarkably, A2597 is one of the few cooling flows toward which
a large column has been detected by 21 cm H~I radio observations
(O'Dea et al.\ 1994a).
O'Dea et al.\ detect a total column of
$\Delta N_H \approx 1 \times 10^{21}$ cm$^{-2}$ in absorption against
the central radio source in the inner 10 kpc of the cD galaxy in A2597.
Our X-ray upper limit is consistent with the detection by
O'Dea et al.\ if the absorber only covers the small region occupied
by the radio source.

We determined the ROSAT PSPC radial X-ray surface brightness profile and
merged this with the previous ROSAT HRI profile for the inner regions
(Paper I).
The profile is not adequately fit by a beta model because of the
central X-ray surface brightness peak associated with the cooling flow.
If the central 108 arcsec in radius are excluded, the fit become
acceptable.
This fit gave $\beta = 0.64^{+0.08}_{-0.03}$
but only an upper limit to the core radius,
$r_{core} < 78$ arcsec (both 90\% confidence).
We de-projected the merged X-ray surface brightness to determine the
profile of gas density as a function of radius.
The hydrostatic equilibrium condition was used to determine the
total gravitational mass as a function of radius in the cluster.
The gas density was integrated to give the gas mass as a function
of radius.
Within a radius of 2 Mpc, we found masses of
$M_{gas} = 1.2 \times 10^{14}$ $M_\odot$ and 
$M_{tot} = 5.6 \times 10^{14}$ $M_\odot$, and
a gas mass fraction of about 21\%
(for a cluster gas temperature of 4.34 keV).
However, our poor knowledge of the spatial variation of the gas
temperature makes the total mass values quite uncertain.

\begin{figure*}[t]
\vskip3.0truein
\includegraphics{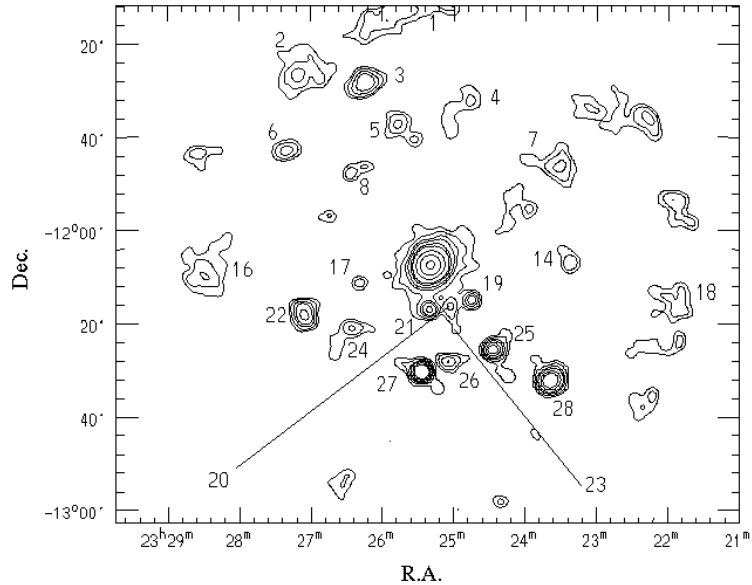}
\caption{A contour plot of the X-ray surface brightness of the
entire field of view of the ROSAT PSPC image of A2597.
The contours are (0.7, 1, 1.5, 2, 3, 4, 10, 30, 100) $\times
10^{-3}$
cts sec$^{-1}$ arcmin$^{-2}$ in the 0.1 -- 2.1 keV ROSAT band.
The image has been adaptively smoothed to a signal-to-noise of
five per smoothing beam.
The individual X-ray sources outside of the central cluster
region are labeled with theirs numbers from
Table~\protect\ref{tab:sources}.
The coordinates are J2000.}
\label{fig:x-ray_whole}
\end{figure*}

%
%
\begin{table*}[b]
\caption[X-Ray Sources]{}
\label{tab:sources}
\begin{center}
\begin{tabular}{lcrllcl}
\multicolumn{7}{c}{\sc Field X-Ray Sources} \cr
\hline \hline
Src.&
\multicolumn{2}{c}{\quad R.A. \hfil (J2000) \hfil Dec. \qquad}&
\multicolumn{2}{c}{Count Rate ($10^{-3}$ s$^{-1}$)}&
$D$&
Possible ID \cr
\cline{4-5}
No.&(h:m:s)&
($^\circ$:$^\prime$:$^{\prime\prime}$)~~&
\multicolumn{1}{c}{PSPC}&
\multicolumn{1}{c}{HRI}
&(arcmin)&\cr
\hline
\phn 1 & 23:25:56.7 & -11:15:01 & 75.3$^a$         &
&
    53 & GD 1148 ? \\
\phn 2 & 23:27:08.5 & -11:27:15 & 14.2$^a$         &
&
    47 & HD 220826 ? \\
\phn 3 & 23:26:14.0 & -11:28:14 & 24.7$^a$         &
&
    41 & \\
\phn 4 & 23:24:46.5 & -11:31:27 & 20.0$^a$         &
&
    37 & \\
\phn 5 & 23:25:46.7 & -11:36:58 & 17.6$^a$         &
&
    30 & HD 220687 \\
\phn 6 & 23:27:20.5 & -11:42:50 & 19.1$^a$         &
&
    37 & \\
\phn 7 & 23:23:30.9 & -11:46:21 & 13.7$^a$         &
&
    34 & \\
\phn 8 & 23:26:27.3 & -11:47:38 & 12.0$^a$         &
&
    25 & \\
\phn 9 & 23:25:25.5 & -12:04:08 & 18.5$\pm$1.9     & \qquad Y
&
\phn 3 & Fluct? \\
    10 & 23:25:15.4 & -12:05:52 & \qquad N         & \phn
2.4$\pm$0.5
    &
\phn 1 & Fluct? \\
    11 & 23:25:15.8 & -12:06:30 & \qquad N         & \phn
6.5$\pm$0.7
    &
\phn 1 & Fluct? \\
    12 & 23:25:21.9 & -12:06:54 & \qquad N         & 13.2$\pm$0.9
    &
\phn 1 & Fluct? \\
    13 & 23:25:17.9 & -12:06:55 & \qquad N         & 25.2$\pm$1.2
    &
\phn 1 & Fluct? \\
    14 & 23:23:22.8 & -12:07:20 & 11.5$^a$         &
    &
    29 & \\
    15 & 23:25:19.1 & -12:08:50 & \qquad Y         & \phn
3.8$\pm$0.6
    &
\phn 2 & Fluct? \\
    16 & 23:28:28.8 & -12:09:07 & 35.3$^a$         &
    &
    45 & \\
    17 & 23:26:17.3 & -12:11:06 & \phn 6.1$\pm$1.2 & \qquad Y
    &
    14 & \\
    18 & 23:22:01.2 & -12:14:43 & 27.4$^a$         &
    &
    50 & HD 220188 \\
    19 & 23:24:44.9 & -12:14:52 & 15.0$\pm$1.6     & \phn
3.3$\pm$0.7
    &
    11 & \\
    20 & 23:25:02.9 & -12:16:13 & \phn 7.4$\pm$1.2 & \phn
2.9$\pm$0.7
    &
    10 & \\
    21 & 23:25:19.9 & -12:17:06 & 16.5$\pm$1.8     & \phn
4.3$\pm$0.7
    &
\phn 9 & \\
    22 & 23:27:04.8 & -12:18:22 & 32.2$^a$         &
    &
    27 & \\
    23 & 23:25:01.2 & -12:18:31 & \phn 4.0$\pm$1.0 & \qquad N
    &
    12 & \\
    24 & 23:26:25.4 & -12:20:46 & 10.4$^a$         &
    &
    21 &  \\
    25 & 23:24:27.3 & -12:25:37 & 24.4$^a$         &
    &
    22 & PHL 5776 ? \\
    26 & 23:25:03.7 & -12:28:45 & 10.2$^a$         &
    &
    22 & PHL 5785 ? \\
    27 & 23:25:27.0 & -12:30:25 & 76.6$^a$         &
    &
    23 & HD 220628 \\
    28 & 23:23:38.5 & -12:32:20 & 60.2$^a$         &
    &
    35 & \\
\hline
\end{tabular}
\end{center}
\tablenotetext{a}{These sources have very uncertain fluxes
because of either their proximity to a rib or their distance
from the center of the observing field.}
\end{table*}

\acknowledgements

C. L. S. was supported in part by NASA ROSAT grants NAG 5--1891,
NAG 5--3308, NASA ASCA grant NAG 5-2526, and NASA Astrophysical
Theory Program grant 5-3057.
B. R. M. received partial support from grant NAS8-39073.
C. L. S. thanks Chris O'Dea for
useful discussions, and for communicating his results on the 21
cm absorption prior to publication.
C. L. S. thanks Zhenping Huang and Jimmy Irwin for helpful advice.
C. L. S. and B. R. M. thank an anonymous referee for a number of
helpful comments and suggestions.


\clearpage

\appendix

\section{APPENDIX --- FIELD X-RAY SOURCES} \label{sec:app_field}

Figure~\ref{fig:x-ray_whole} gives a contour plot of the
X-ray image of essentially the entire field of view in the
photon energy band 0.1 -- 2.1 keV (PI channels 11--201).
This image was smoothed with an adaptive kernel routine which
convolved the image such that each smoothing beam had a minimum
signal-to-noise ratio of five; the minimum smoothing length was
defined by the PSPC point-spread-function
(Huang \& Sarazin 1996).
The cluster is at the center of the image.
Regions of low exposure near the edge of the image were eliminated.
Some of the features at the circular edge of the field of view of
the telescope are enhanced by the large vignetting and exposure
corrections there, and may be spurious.

Maximum-likelihood and local detection algorithms were used to
detect point sources in the PSPC and HRI images.
A detection criterion of 5$\sigma$ was adopted.
Table~\ref{tab:sources} gives a list of the sources detected
in either the PSPC or the HRI (excluding the extended source
associated with A2597 itself).
The sources outside of the crowded central region are also labeled
in Figure~\ref{fig:x-ray_whole}.
For each source, the table gives its centroid position, its
count rate (corrected for background and vignetting) in the
PSPC and/or HRI, its projected distance $D$ from the center of
the X-ray image, and a comment on the identification.
The sources at large distances from the center of the field
or near the detector ribs have poorly determined fluxes and
positions.
None of the sources was clearly extended;
however, for sources far from the center of the field where the
instrumental point-spread-function is very broad, the upper
limits on their size are large.

Only three sources (Src.\ 19, 20, and 21)
were detected clearly in both the HRI and PSPC.
Because the HRI has a smaller field of view, sources at large
distances from the center of A2597 were only visible in the
PSPC.
For the region of overlapping coverage, we examined the HRI image
at the position of the PSPC sources (and vice versa).
We indicate in the table if an enhancement was detected at the
position of the source in the other detector (Y) or not (N).
A Y indicates that an enhancement is seen, but that it did not
satisfy the 5$\sigma$ detection criteria, either because
if was too faint, or resolved out, or possibly varied with time.
A number of sources were observed with the HRI near the cluster
core which may well just be fluctuations in the surface
brightness
of the cluster emission;
these are noted in the Possible ID column of the Table.
Some other possible identifications are noted.
Based on the positional errors and the X-ray to optical
flux ratios, only the identifications of Srcs.\ 5, 18, and 27
seem likely to be correct.


\end{document}